\documentclass[aps,prd,nofootinbib,floatfix,amsmath,amssymb,colorBG,spanish]{revtex4}
\usepackage{amssymb}
\usepackage{graphicx}
\usepackage{color}
\usepackage[tight,TABTOPCAP]{subfigure}
\usepackage{epstopdf}
\begin{document}
\makeatletter

%Feynman slash
\newbox\slashbox \setbox\slashbox=\hbox{$/$}
\newbox\Slashbox \setbox\Slashbox=\hbox{$/$}
\def\pFMslash#1{\setbox\@tempboxa=\hbox{$#1$}
  \@tempdima=0.5\wd\slashbox \advance\@tempdima 0.5\wd\@tempboxa
  \copy\slashbox \kern-\@tempdima \box\@tempboxa}
\def\pFMSlash#1{\setbox\@tempboxa=\hbox{$#1$}
  \@tempdima=0.5\wd\Slashbox \advance\@tempdima 0.5\wd\@tempboxa
  \copy\Slashbox \kern-\@tempdima \box\@tempboxa}
\def\FMslash{\protect\pFMslash}
\def\FMSlash{\protect\pFMSlash}
\def\miss#1{\ifmmode{/\mkern-11mu #1}\else{${/\mkern-11mu #1}$}\fi}
%%%% Uso:  \pFMSlash{p}
\makeatother

%%%%%%%%%%%%%%%%%%%%%%%%%%%%%%%%%%%%%%%%%%%%%%%%%%%%%%%%%%%%%%%%%

\title{The $Z_H \to \gamma H$ decay in the Littlest Higgs Model}

\author{J. I. Aranda$^{(a)}$, I. Cort\'es-Maldonado$^{(b,c)}$, F. Ram\'\i rez-Zavaleta$^{(a)}$, E. S. Tututi$^{(a)}$}

\address{$^{(a)}$Facultad de Ciencias F\'\i sico Matem\' aticas, Universidad Michoacana de San Nicol\' as de Hidalgo, Avenida Francisco J. M\' ujica S/N, 58060, Morelia, Michoac\'an, M\' exico. \\
$^{(b)}$Departamento de F\'{\i}sica, CINVESTAV IPN, Apartado Postal 14-740, 07000, M\'exico D. F., M\'exico. \\
$^{(c)}$C\'atedras Consejo Nacional de Ciencia y Tecnolog\'ia, M\'exico.}

\begin{abstract}
We present the calculation of the $Z_H \to \gamma H$ decay in the context of the Littlest Higgs model at one-loop level. Our calculations include the contributions of fermions, scalars and gauge bosons in accordance with the most recent experimental constraints on the parameters space of the model. We find branching ratios of the order of $10^{-5}$ for the energy scale $f=2,3,4$ TeV on the $0.1<c<0.9$ region. In order to provide a complementary study we calculated the production cross section of the $Z_H$ boson in $pp$ collisions at Large Hadron Collider with a center of mass energy of 14 TeV. By using the integrated luminosity projected for the Large Hadron Collider in the last stage of operation, we estimated the number of events for this process. Moreover, we analyze the SM background for the Higgs-photon associated production and found that the $pp\to Z_H X \to \gamma H$ production is above the SM background.
\end{abstract}

\pacs{12.60.Cn, 12.60.Fr, 14.70.-e}
%\keywords{Extended models, Heavy bosons}

\maketitle
\section{Introduction}
\label{SecI}
Alternative formulations for the study of electroweak symmetry breaking that have the property of canceling quadratic divergences are the so called little Higgs models (LHM)~\cite{LHMs,LHM1}. These models are based on dimensional deconstruction~\cite{Dec1,Dec2}, where the quadratic divergence induced at the one-loop level by the Standard Model gauge bosons are canceled via the quadratic divergence introduced by heavy gauge bosons at the same perturbative level. Also, it is proposed the existence of heavy-mass fermions interacting with the Higgs Field in such a way that the one-loop quadratic divergence induced in the Yukawa sector of the Standard Model (SM) due to top quark coupling with the Higgs boson is canceled~\cite{LHM1,Han:2003wu}. Furthermore, the Higgs fields acquire mass becoming pseudo-Goldstone bosons via an approximate global symmetry breaking, where a massless Higgs appears. Quadratically divergent corrections to the Higgs mass arise at loop level, therefore, this naturally ensure a light Higgs.

As far as the littlest Higgs (LTHM) is concerned, a remarkable feature is that there is no new degrees of freedom beyond the SM below TeV scale. Moreover, above few TeV's the LTHM needs a very small new degrees of freedom to stabilize the Higgs boson mass. At the TeV energy scale, the arising new particles are a set of four gauge bosons with the same quantum numbers as the electroweak SM gauge bosons, namely, $A_H$, $Z_H$, and $W^{\pm}_H$, an exotic quark with the same charge as the top quark, and a scalar triplet~\cite{LHM1}. The construction details of the model can be found in Refs.~\cite{LHMs,LHM1,Han:2003wu}. In general, these extensions of the SM predict new particles emerging at the TeV scale and the new physics that could appear at these energies that soon will be tested at the Large Hadron Collider (LHC)~\cite{LHCexp}. In particular, little Higgs models predict the existence of a new neutral massive gauge boson, known as $Z_H$, which could offer another theoretical framework to justify the experimental scrutiny about the possible existence of heavy-mass (at the TeV scale) particles like the $Z$ gauge boson of the SM. On the other hand, there are several models that predict the existence of a neutral massive gauge boson, identified as $Z^\prime$ gauge boson, such as the 331 model~\cite{331} or grand unified models~\cite{GUT}. These type of particles are under exhaustive search at the LHC~\cite{Zpatlas-mass,Zpcms-mass}, where the ATLAS and CMS collaborations have imposed experimental bounds over the mass of a new particle related to $Z^\prime$ gauge boson, their results indicate that the mass of the $Z^\prime$ gauge boson must be greater than $2.49$ TeV and $2.59$ TeV, respectively.

In this work we are interested in the physics of the $Z_H$ gauge boson, specifically, the main concern of this paper is to study the $Z_H\to H\gamma$ decay in the context of the linearized theory of the littlest Higgs model~\cite{Han:2003wu}. The relevance of this process brings the possibility of testing the LTHM, since the parameters space has been severely constrained by the Higgs discovery channels and electroweak precision observables~\cite{Reuter}. Another remarkable feature of the $Z_H\to H\gamma$ decay consists in the fact the SM background is naturally suppressed, namely, the $pp\to H\gamma$ reaction, since it is highly suppressed for its electroweak origin~\cite{Abbasabadi:1997zr,Passarino:2013nka}. Thus, the Higgs-photon associated production opens a new window to test the gauge sector of the SM and Higgs physics~\cite{Toscano,Martinez,Aad:2010sp,Aad:2013zba}. To support our analysis, it is calculated the Higgs-photon associated production at LHC coming from SM background, by using current kinematical cuts employed by ATLAS Collaboration~\cite{Aad:2010sp,Aad:2013zba}. Previous studies on the $Z^\prime\to H\gamma$ decay have been performed in the context of left-right symmetric models~\cite{Toscano}, where the associated branching ratio is estimated, however, the used parameters such as the $m_{Z^\prime}$ mass are below the present bounds established by the experimental measurements~\cite{Zpatlas-mass,Zpcms-mass}.

%we observed that SM-background calculations for the $pp\to H\gamma$ are below the calculated LTHM signal corresponding to the $Z_H\to H\gamma$ decay.

The paper is organized as follows. In Section~\ref{SecMF}, we briefly describe the theoretical framework of the LTHM. In Sec.~\ref{SecD}, we outline the analytical results for the $Z_H\to H\gamma$ decay in the LTHM. In Sec.~\ref{NR}, it is presented the numerical analysis. Finally, the conclusions appear in Sec.~\ref{CON}.

\section{Model framework}
\label{SecMF}

The littlest Higgs model is based on a nonlinear sigma model with $SU(5)$ global symmetry and the gauged subgroup $[SU(2)_1\otimes U(1)_1]\otimes [SU(2)_2\otimes U(1)_2]$~\cite{LHM1,Han:2003wu}. The global symmetry of the $SU(5)$ group is spontaneously broken down $SU(5)\to SO(5)$ at the energy scale $f$, where $f$ is constrained to be of the order of $2\textendash4$ TeV~\cite{Reuter}. Simultaneously, the $[SU(2)_1\otimes U(1)_1]\otimes [SU(2)_2\otimes U(1)_2]$ group is also broken to its subgroup $SU_L(2)\otimes U_Y(1)$, which results to be the SM electroweak gauge group. The global symmetry breaking pattern leaves 14 Goldstone bosons which transform under the $SU_L(2)\otimes U_Y(1)$ group as a real singlet $\mathbf{1}_0$, a real triplet $\mathbf{3}_0$, a complex doublet $\mathbf{2}_{\pm \frac{1}{2}}$, and a complex triplet $\mathbf{3}_{\pm 1}$~\cite{LHM1,Han:2003wu}. The spontaneous global symmetry breaking of the $SU(5)$ group is generated by the vacuum expectation value (VEV) of the $\Sigma$ field, denoted as $\Sigma_0$~\cite{Han:2003wu}, at the scale $f$, which is parametrized by
\begin{equation}
\Sigma=e^{i\Pi/f}\Sigma_0e^{i\Pi^T/f},
\end{equation}
with
\begin{equation}
\Sigma_0=\left(\begin{array}{ccc}
\mathbf{0}_{2\times 2} & \mathbf{0}_{2\times 1} & \mathbf{1}_{2\times 2}\\
\mathbf{0}_{1\times 2} & 1 & \mathbf{0}_{1\times 2}\\
\mathbf{1}_{2\times 2} & \mathbf{0}_{2\times 1} & \mathbf{0}_{2\times 2}
\end{array}\right)
\end{equation}
and $\Pi$ being the Goldstone boson matrix given by
\begin{equation}
\Pi=\left(\begin{array}{ccc}
\mathbf{0}_{2\times 2} & h^\dagger/\sqrt{2} & \phi^\dagger\\
h/\sqrt{2} & 0 & h^\ast/\sqrt{2}\\
\phi & h^T/\sqrt{2} & \mathbf{0}_{2\times 2}
\end{array}\right).
\end{equation}
Here, $h$ is a doublet and $\phi$ is a triplet under the $SU_L(2)\otimes U_Y(1)$ SM gauge group~\cite{Han:2003wu}. By the spontaneous symmetry breaking (SSB), both the real singlet and the real triplet are absorbed by the longitudinal components of the gauge bosons at the energy scale $f$. At this scale, the complex doublet and the complex triplet remain massless. The complex triplet acquires a mass of the order of $f$ by means of the Coleman-Weinberg type potential when the global symmetry of the group $SO(5)$ breaks down. The complex doublet is identified as the SM Higgs field.

The effective Lagrangian invariant under the $[SU(2)_1\otimes U(1)_1]\otimes [SU(2)_2\otimes U(1)_2]$ group is~\cite{Han:2003wu}
\begin{equation}
\mathcal{L}_{LTHM}=\mathcal{L}_G+\mathcal{L}_F+\mathcal{L}_\Sigma+\mathcal{L}_Y-V_{CW},
\end{equation}
where $\mathcal{L}_G$ represents the gauge bosons kinetic contributions, $\mathcal{L}_F$ the fermion kinetic contributions, $\mathcal{L}_\Sigma$ the non-linear sigma model contributions of the LTHM, $\mathcal{L}_Y$ the Yukawa couplings of fermions and pseudo-Goldstone bosons, and the last term symbolizes the Coleman-Weinberg potential.

The standard form of the Lagrangian of the non-linear sigma model is
\begin{equation}
\mathcal{L}_\Sigma=\frac{f^2}{8}\mathrm{tr}\left|\mathcal{D}_\mu\Sigma\right|^2,
\end{equation}
where the covariant derivative is written as
\begin{equation}
\mathcal{D}_\mu\Sigma=\partial_\mu\Sigma-i\sum\limits_{j=1}^{2}\left[g_j \sum\limits_{a=1}^{3}W_{\mu j}^{a}\left(Q_j^a\Sigma+\Sigma Q_j^{a T}\right)+g^\prime_j B_{\mu j}\left(Y_j\Sigma+\Sigma Y_j^{T}\right)\right].
\end{equation}
Here, $W_{\mu j}^{a}$ are the $SU(2)$ gauge fields, $B_{\mu j}$ are the $U(1)$ gauge fields, $Q_j^a$ are the $SU(2)$ gauge group generators, $Y_j$ are the $U(1)$ gauge group generators, $g_j$ are the coupling constants of the $SU(2)$ group, and $g_j^\prime$ are the coupling constants of the $U(1)$ group~\cite{Han:2003wu}. After SSB around $\Sigma_0$, it is generated the mass eigenstates of order $f$ for the gauge bosons~\cite{Han:2003wu}
\begin{eqnarray}
W^\prime_\mu&=&-cW_{\mu 1}+sW_{\mu 2},\\
B^\prime_\mu&=&-c^\prime \mathcal{B}_{\mu 1}+s^\prime \mathcal{B}_{\mu 2},\\
W_\mu&=& sW_{\mu 1}+cW_{\mu 2},\\
B_\mu&=& s^\prime \mathcal{B}_{\mu 1}+c^\prime \mathcal{B}_{\mu 2},
\end{eqnarray}
where $W_{\mu j}\equiv\sum\limits_{a=1}^{3}W^a_{\mu j}Q^a_j$ and $\mathcal{B}_{\mu j}\equiv B_{\mu j}Y_j$ for $j=1,2$; $c=g_1/\sqrt{g_1^2+g_2^2}$, $c^\prime=g^\prime_1/\sqrt{g_1^{\prime 2}+g_2^{\prime 2}}$, $s=g_2/\sqrt{g_1^2+g_2^2}$, and $s^\prime=g^\prime_2/\sqrt{g_1^{\prime 2}+g_2^{\prime 2}}$. Notice that $\Sigma$ field has been expanded around $\Sigma_0$ holding dominant terms in $\mathcal{L}_\Sigma$~\cite{Han:2003wu}. At this stage of SSB the $B_\mu$ and $W_\mu$ fields remain massless.

The SSB at the Fermi scale provides mass to the SM gauge bosons ($B$ and $W$) and induces mixing between heavy and light gauge bosons. The arising masses at the leading order (neglecting terms of order $\mathcal{O}\left(\frac{v^2}{f^2}\right)$, with $v$ being the vacuum expectation value at the Fermi scale) are~\cite{Reuter}
\begin{eqnarray}
m_{Z_H}&=&\frac{gf}{2sc},\\
m_{A_H}&=&\frac{g^\prime f}{2\sqrt{5}s^\prime c^\prime},\\
m_{W_H}&=&\frac{gf}{2sc}.
\end{eqnarray}
As it is known $c=m_{W_{H}}/m_{Z_{H}}$ and takes the value equals to one at the leading order, we may assume that the $c$ parameter ranges from $0.1$ to $0.9$~\cite{Han:2003wu}, in order to have values for the masses of the weak gauge bosons not very different, as it occurs in the electroweak sector of the SM.

The LTHM incorporate new heavy fermions which couple to Higgs field in a such way that the quadratic divergence of the top quark is canceled~\cite{LHM1,Han:2003wu}. In particular, this model introduces a new set of heavy fermions arranged as a vector-like pair ($\tilde{t},\tilde{t}^{\prime c}$) with quantum numbers $(\mathbf{3},\mathbf{1})_{Y_i}$ and $(\bar{\mathbf{3}},\mathbf{1})_{-Y_i}$, respectively. The new Yukawa interactions are proposed to be
\begin{equation}
\mathcal{L}_Y=\frac{1}{2}\lambda_1\, f\,\epsilon_{ijk}\epsilon_{xy}\,\chi_i\,\Sigma_{jx}\,\Sigma_{ky}u_3^{\prime c}+\lambda_2\,f\,\tilde{t}\tilde{t}^{\prime c}+\mathrm{H.c.},
\end{equation}
where $\chi_i=(b_3,t_3,\tilde{t})$; $\epsilon_{ijk}$ and $\epsilon_{xy}$ are antisymmetric tensors for $i,j,k=1,2,3$ and $x,y=4,5$~\cite{LHM1}. Here, $\lambda_1$ and $\lambda_2$ are free parameters, where the $\lambda_2$ parameter can be fixed such that, for given $(f,\lambda_1)$, the top quark mass adjust to its experimental value~\cite{Reuter}.

Expanding the $\Sigma$ field and retaining terms up to $\mathcal{O}(v^2/f^2)$ after diagonalizing the mass matrix, it can be obtained the mass states $t_L$, $t^c_R$, $T_L$, and $T^c_R$, which correspond to SM top quark and the heavy top quark, respectively~\cite{Han:2003wu,Reuter}.

The explicit remaining terms of the Lagrangian $\mathcal{L}_{LTHM}$ as well as the complete set of new Feynman rules can be found in Ref.~\cite{Han:2003wu}.

\section{Decay $Z_H\to \gamma H$}
\label{SecD}

\begin{figure}[!ht]
\begin{center}
\includegraphics[scale=0.7]{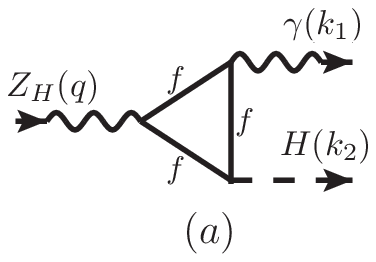}\\
\includegraphics[scale=0.7]{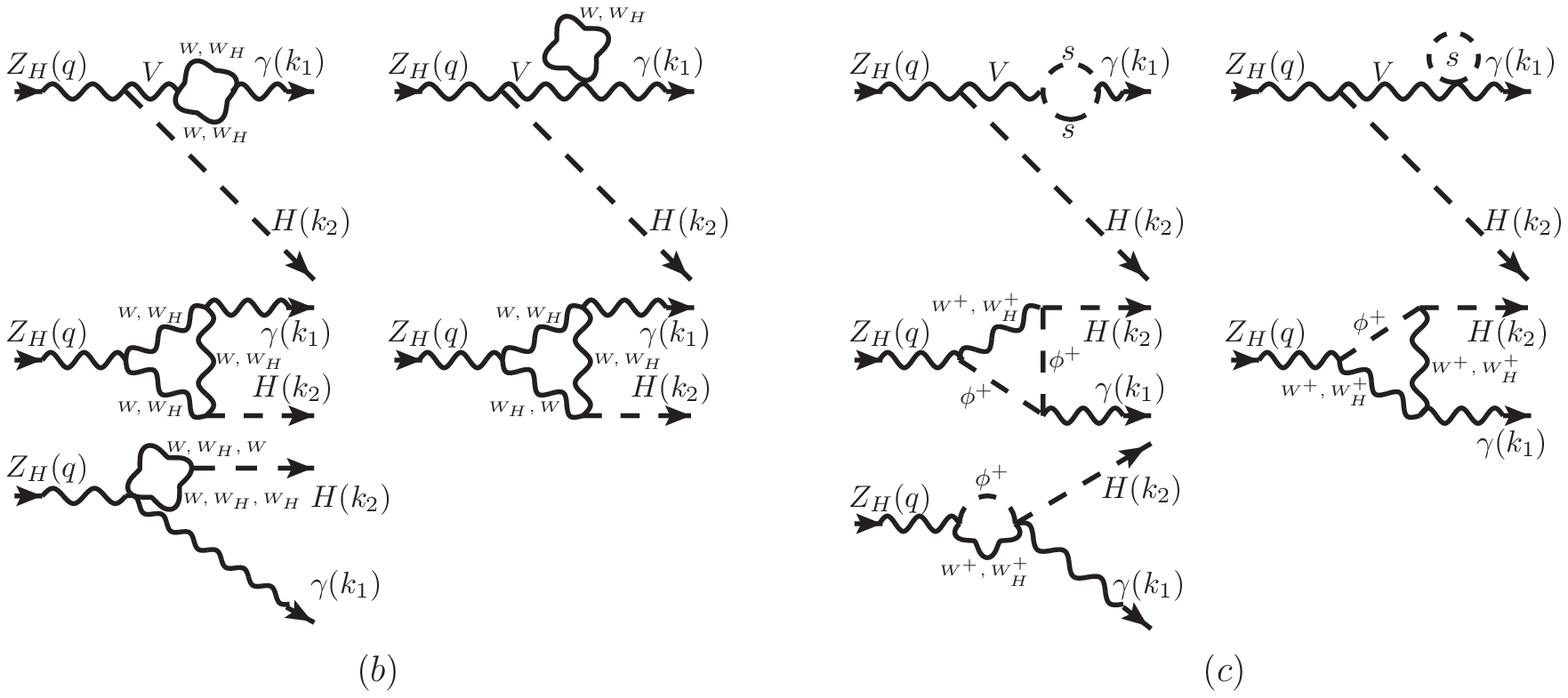}
\caption{Feynman diagrams contributing to the $Z_H \to \gamma H$ decay at one-loop level. Here, $f=u, d, c, s, t, b, e, \mu, \tau, T$, $V=Z, Z_H, A_H$ and $s=\phi^+, \phi^{++}$.}
\label{feyn-diagrams-contrib}
\end{center}
\end{figure}

We now turn our attention to obtain the analytical expression for the amplitude and decay width of the $Z_H\to \gamma H$ process. The amplitude was calculated in the unitary gauge. In Fig. \ref{feyn-diagrams-contrib} we show the contributions at one-loop level to the $Z_H\to \gamma H$ coming from fermions, gauge boson and scalars. In the fermion loops we include SM and LTHM fermion contributions. The analysis is based on three sets of Feynman diagrams: the set (a) contains the triangle loop contributions mediated by fermions; the set (b) includes triangle and bubble loop contributions mediated by SM charged gauge bosons and new heavy charged gauge bosons, where also it is include the mixing effects between these two types of charged gauge bosons; for the set (c), we take into account the bubble loop contributions induced by scalars and scalars plus gauge bosons, together with triangle loop contributions mediated by gauge bosons and scalars.

The respective decay amplitude is given by
\begin{equation}
\mathcal{M}(Z_H \to \gamma H)=\mathcal{M}_T^{\mu \nu}\epsilon_\mu(q)\epsilon_\nu(k_1),
\end{equation}
where $\mathcal{M}_T^{\mu \nu}=\mathcal{M}_f^{\mu \nu}+\mathcal{M}_G^{\mu \nu} + \mathcal{M}_S^{\mu \nu}$. Here, $\mathcal{M}_f^{\mu \nu}$ represents the contribution of the set (a), $\mathcal{M}_G^{\mu \nu}$ contains the contribution of the set (b), and $\mathcal{M}_S^{\mu \nu}$ includes the contribution of the set (c). Moreover, $\epsilon_\mu(q)$ and $\epsilon_\nu(k_1)$ are the polarization vectors associated to $Z_H$ boson and photon, respectively. After tedious algebraic manipulations we can write down a generic expression for the total decay amplitude as follows
\begin{equation}
\mathcal{M}_T^{\mu \nu}=\mathcal{A}_T g^{\mu \nu}+ \mathcal{B}_T \hat k_1^\mu \hat q^\nu+\,\mathcal{C}_T\varepsilon^{\mu\nu\alpha\beta}k_{1\,\alpha}q_{\beta},
\label{ampW}
\end{equation}
where $\hat k_1=k_1/m_{Z_H}$ and $\hat q=q/m_{Z_H}$. The $\mathcal{A}_T$, $\mathcal{B}_T$ and $\mathcal{C}_T$ coefficients are in terms of Passarino-Veltman scalar functions. In specific, $\mathcal{A}_T=\sum\limits_{f}^{}\mathcal{A}_f+\sum\limits_{i=1}^{3}\mathcal{A}_{G_{i}}+\sum\limits_{i=1}^{2}\mathcal{A}_{S_{i}}$, $\mathcal{B}_T=\sum\limits_{f}^{}\mathcal{B}_f+\sum\limits_{i=1}^{3}\mathcal{B}_{G_{i}}+\sum\limits_{i=1}^{2}\mathcal{B}_{S_{i}}$ and $\mathcal{C}_T=\sum\limits_{f}^{}\mathcal{C}_f$. Here, $f$ runs over all charged fermions, $G_i$ represents charged gauge bosons ($W$, $W_H$), and $S_i$ symbolizes charged scalars
($\phi^+$, $\phi^-$, $\phi^{++}$, and $\phi^{--}$). We found that the total contribution arising from tadpole and self-energies diagrams vanishes.

The explicit form for the $\mathcal{A}_f$, $\mathcal{A}_{G_{i}}$, $\mathcal{A}_{S_{i}}$, $\mathcal{B}_f$, $\mathcal{B}_{G_{i}}$, $\mathcal{B}_{S_{i}}$ and $\mathcal{C}$ coefficients are presented below
\begin{eqnarray}
\mathcal{A}_f&=&\frac{g^2}{8 m_W}\frac{m^2_{Z_H}\,\sqrt{y_f}\,(h_L+h_R)}{y_H-1}
\left(2(B_a-B_b)\right.\nonumber\\
&&+\left.(y_H-1)(C_a(4y_f-y_H+1)+2) \right),\\
\mathcal{B}_f&=&\frac{2}{(y_H-1)}\,\mathcal{A}_{f},
\end{eqnarray}
where $B_a\equiv B_0(m_H^2,m_f^2,m_f^2)$, $B_b\equiv B_0(m_{Z_H}^2,m_f^2,m_f^2)$ and $C_a\equiv m_{Z_H}^2 C_0(m_H^2,m_{Z_H}^2,m_f^2,m_f^2,m_f^2)$ are the known Passarino-Veltman scalar functions~\cite{Passarino}. Also, we used $y_{f}=m_f^2/m_{Z_H}^2$ and $y_{H}=m_H^2/m_{Z_H}^2$. Notice that for this particular process $h_R=0$ and $h_L=g\,c\,T^3/s$, where, as usual $T^3$ represents the third component of isospin being $T^3=1\,(-1)$ for fermions up (down) type~\cite{Han:2003wu}. The $\mathcal{A}_{G_{i}}$ and $\mathcal{B}_{G_{i}}$ coefficients contain the contributions of $W$ and $W_H$ bosons as follows
\begin{eqnarray}
\mathcal{A}_{G_{1}} &=& C_{G_{1}} \frac{1}{4 (y_H-1)y_W^2}\Big( (B_{G_{1a}}-B_{G_{1b}}) (y_H (1-2 y_W)+2 (1-6 y_W) y_W)\nonumber\\ &-& 2 C_{G_{1a}}\, y_W \left(y_H^2
   (1-6 y_W)+3 y_H \left(4 y_W^2+4 y_W-1\right)-12 y_W^2-6 y_W+2\right)\nonumber\\ &+& y_H^2 (1-2 y_W)+y_H \left(-12 y_W^2+4 y_W-1\right)+2 y_W (6 y_W-1)\Big),\\
\mathcal{B}_{G_{1}} &=& \frac{2}{(y_H-1)}\,\mathcal{A}_{G_{1}},
\end{eqnarray}
where $B_{G_{1a}}\equiv B_0(m_H^2,m_W^2,m_W^2)$, $B_{G_{1b}}\equiv B_0(m_{Z_H}^2,m_W^2,m_W^2)$, and $C_{G_{1a}}\equiv m_{Z_H}^2 C_0(m_H^2,m_{Z_H}^2,m_W^2,m_W^2,m_W^2)$. Moreover, $y_W=m_W^2/m_{Z_H}^2$ and $C_{G_1}=-\frac{1}{2 f^2}
\Big(c g^4 s \Big(c^2-s^2\Big) s_W v^3\Big)$. The $\mathcal{A}_{G_{2}}$ and $\mathcal{B}_{G_{2}}$ coefficients can be obtained from $\mathcal{A}_{G_{1}}$ and $\mathcal{B}_{G_{1}}$ by replacing: $m_W\to m_{W_H}$ and $C_{G_1}\to C_{G_2}$, where $C_{G_2}=-\frac{1}{2 c s}
\Big(g^4 s_W v \Big(c^2-s^2\Big)\Big)$. The $\mathcal{A}_{G_{3}}$ and $\mathcal{B}_{G_{3}}$ coefficients are given by
\begin{eqnarray}
\mathcal{A}_{G_3} &=& C_{G_3} \frac{1}{2 (y_H-1) y_W y_{W_H}} \Big((B_{G_{3a}}-B_{G_{3b}}) \Big[-y_{W_H} (y_H+10 y_W-1)\nonumber\\
&-&(y_W-1)(y_H+y_W)-y_{W_H}^2\Big] - C_{G_{3a}} (y_H-1) y_W \Big[y_H (1-y_W-5y_{W_H})\nonumber\\
&+& y_W^2+10 y_W y_{W_H}+y_W+y_{W_H}^2+5 y_{W_H}-2\Big] - C_{G_{3b}} (y_H-1)y_{W_H}\nonumber\\
&\times&\Big[y_H (1-5 y_W-y_{W_H})+y_W^2+5 y_W (2
   y_{W_H}+1)+y_{W_H}^2+y_{W_H}-2\Big]\nonumber\\
   &-& (y_H-1) \left(y_{W_H} (y_H+10 y_W-1)+(y_W-1)(y_H+y_W)+y_{W_H}^2\right) \Big)\label{intGa}\\
\mathcal{B}_{G_3} &=& \frac{2}{(y_H-1)}\,\mathcal{A}_{G_{3}}\label{intGb},
\end{eqnarray}
where $B_{G_{3a}}\equiv B_0(m_H^2,m_W^2,m_{W_H}^2)$, $B_{G_{3b}}\equiv B_0(m_{Z_H}^2,m_W^2,m_{W_H}^2)$, $C_{G_{3a}}\equiv m_{Z_H}^2 C_0(m_H^2,m_{Z_H}^2,m_W^2,m_{W_H}^2,m_W^2)$, and\\ $C_{G_{3b}}\equiv m_{Z_H}^2 C_0(m_H^2,m_{Z_H}^2,m_{W_H}^2,m_W^2,m_{W_H}^2)$. In addition, $y_{W_H}=m_{W_H}^2/m_{Z_H}^2$ and $C_{G_3}=g^4 s_W\, v\,(c^2-s^2)/4 c s$. Notice that Eqs.~(\ref{intGa}) and (\ref{intGb}) reflect the mixing between $W$ and $W_H$ gauge bosons. The $\mathcal{A}_{S_{1}}$ and $\mathcal{B}_{S_{1}}$ coefficients are
\begin{eqnarray}
\mathcal{A}_{S_{1}} &=& C_{S_{1}} \frac{1}{(y_H-1) y_W} \Big( (B_{S_{1a}}-B_{S_{1b}}) (y_H+y_W-y_\phi)\nonumber \\
              &+& C_{S_{1a}} (y_H-1) y_\phi (y_H+y_W-y_\phi) - C_{S_{1b}} (y_H-1) y_W (y_H y_W+y_\phi -2)\nonumber\\
              &+&(y_H-1)(y_H+y_W-y_\phi) \Big),\\
\mathcal{B}_{S_{1}} &=&\frac{2}{(y_H-1)}\,\mathcal{A}_{S_{1}},
\end{eqnarray}
where $B_{S_{1a}}\equiv B_0(m_H^2,m_W^2,m_{\phi}^2)$, $B_{S_{1b}}\equiv B_0(m_{Z_H}^2,m_W^2,m_{\phi}^2)$, $C_{S_{1a}}\equiv m_{Z_H}^2 C_0(m_H^2,m_{Z_H}^2,m_W^2,m_{\phi}^2,m_W^2)$, and $C_{S_{1b}}\equiv m_{Z_H}^2 C_0(m_H^2,m_{Z_H}^2,m_{\phi}^2,m_W^2,m_{\phi}^2)$. Here, $y_\phi=m_\phi^2/m_{Z_H}^2$ and $C_{S_1}=e g^3 \left(c^2 - s^2\right) v^{\prime 2}/(2 c s v)$. The $\mathcal{A}_{S_{2}}$ and $\mathcal{B}_{S_{2}}$ coefficients can be get by replacing: $m_W\to m_{W_H}$ and $C_{S_1}\to C_{S_2}$, where $C_{S_2}=e g^3 (c^2 - s^2) \left(c^4 + s^4\right) v^{\prime 2}/(4 c^3 s^3 v)$. The form factor $\mathcal{C}_f$ is given by
\begin{equation}
\mathcal{C}_f=\frac{ig^2}{8m_W} \frac{2 (h_L-h_R)y_f}{y_H-1} (2(B_{a}-B_{b})+C_{a} (y_H-1)).
\end{equation}
\noindent Therefore, the expression for the decay amplitude of the $Z_H\to \gamma H$ can be written as
\begin{equation}
\mathcal{M}(Z_H \to \gamma H)=\Big[\mathcal{A}_T \Big(g^{\mu \nu}+ \frac{2}{y_H-1} \hat k_1^\mu \hat q^\nu \Big)+\,\mathcal{C}_T\varepsilon^{\mu\nu\alpha\beta}k_{1\,\alpha}q_{\beta}\Big]\epsilon_\mu(q)\epsilon_\nu(k_1).\label{amplitudef}
\end{equation}

Finally, the decay width for the process reads as
\begin{equation}
\Gamma(Z_H \to \gamma H)= \frac{(1-y_H)[4\mathcal{A}_T^2+{\mathcal{C}_T}^2\,m^4_{Z_H}(y_H-1)^2]}{96\, \pi\, m_{Z_H}}.
\end{equation}

It should be recalled that all the $\mathcal{A}_i$ and $\mathcal{C}_f$ terms in $\Gamma(Z_H \to \gamma H)$ are free of ultraviolet divergences and the Lorentz structure in Eq.~(\ref{amplitudef}) satisfies the Ward identity $k_{1\nu}\mathcal{M}_T^{\mu\nu}=0$.

% % % % % % % % % % % % % % % % % % % % % % % % % % % % % % % % %
% % % % % % % % % % % % % % % % % % % % % % % % % % % % % % % % %
\section{Numerical results}
\label{NR}

\subsection{Production of extra neutral $Z_H$ boson}
\begin{figure}[htb!]
\centering
\subfigure[]{\includegraphics[scale=0.66]{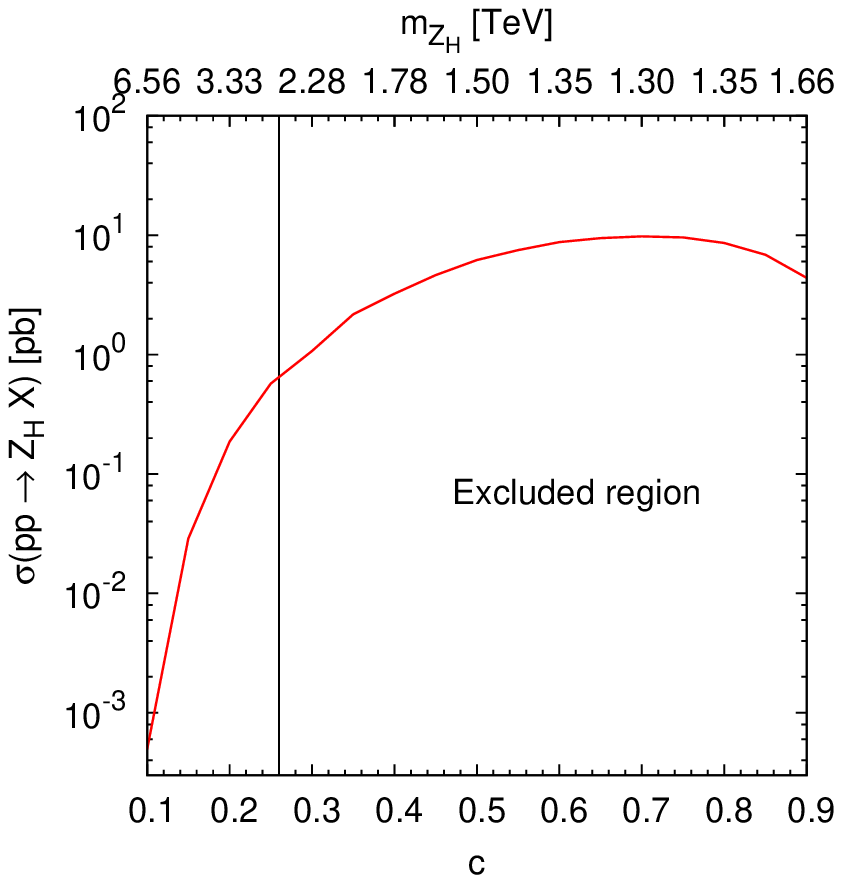}}\;
\subfigure[]{\includegraphics[scale=0.66]{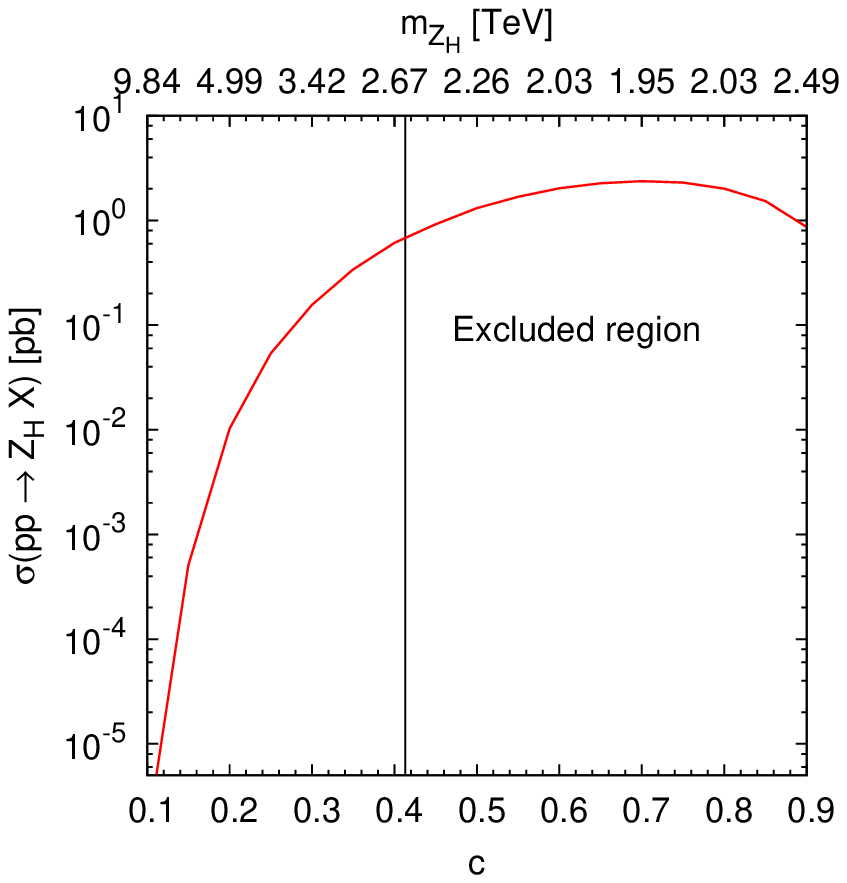}}
\subfigure[]{\includegraphics[scale=0.66]{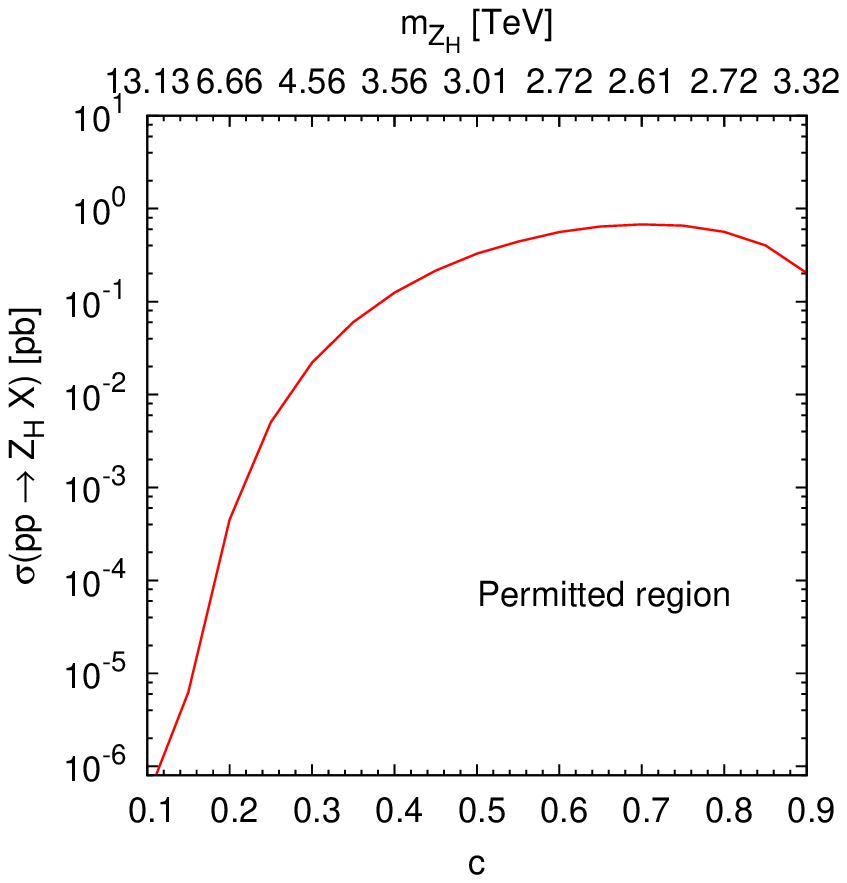}}
\caption{Production cross section of the $Z_H$ boson at LHC.
\label{ZHprod} (a) For $f=2$ TeV. (b) For $f=3$ TeV. (c) For $f=4$ TeV.}
\end{figure}

In this part of our work, we present the production cross section of the extra neutral $Z_H$ gauge boson at LHC in the context of the LTHM, where it is assumed a search for a mass resonance in the dilepton channel $e^+e^-$~\cite{ATLAS-2004}. Here, we used version 2.1.1 of the WHIZARD event generator, which is a program designed for the calculation of multi-particle scattering cross sections and simulated event samples \cite{Kilian:2007gr} to perform our calculations. As a test of the WHIZARD package, we carried out the calculation of $\sigma(pp \to Z_H X)$ cross section as a function of $m_{Z_H}$ for $c=\pi/4$ and our results coincided with previous ones reported in Ref.~\cite{Han:2003wu}; to do that we employed CTEQ5 parton distribution function~\cite{PDF}. We simulate $pp$ collisions with a center of mass energy of 14 TeV. In Fig. \ref{ZHprod}, it can be appreciated $\sigma(pp \to Z_H X)$ as a function of the $Z_H$ boson mass throughout the interval $1.6$ TeV$<m_{Z_H}<13.13$ TeV, where we have employed three different values for $f$, namely, $2,3,4$ TeV. It should be noted that the mass range for $Z_H$ comes from the discrete choice of the energy scale $f$ and allowed values for the $c$ parameter, which will be justified below.

\subsection{Branching ratio for the $Z_H \to \gamma H$ decay}
Let us analyze the branching ratio behavior of the $Z_H \to \gamma H$ decay. Recall that the $m_{Z_H}$ is a function of two model-dependent parameters $c$ and $f$, where $c$ and $f$ are the mixing angle of the $SU(2)_2 \times U(1)_2$ extended gauge group and the energy scale at which the $SU(5)$ gauge group breaks into $SO(5)$ group, respectively. It is known that experimental data constrain the symmetry breaking scale to be in the interval $2\,\,\mathrm{TeV}<f<4\,\,\rm{TeV}$, for $c^\prime=1/\sqrt{2}$ and $c$ between $[0.1,0.995]$~\cite{Reuter}. To make predictions, we will take three distinct values for $f$, namely, $f=2,3,4$ TeV and will carry out an exhaustive study at the $c$ parameter region given above. This analysis will provide us crucial information to test the experimental possibility for $Z_H \to \gamma H$ decay in the LTHM context. Moreover, it should be recalled that the recent results reported by ATLAS and CMS collaborations, established lower mass limits for a new neutral massive gauge boson, identified as $Z^\prime$. ATLAS collaboration reports that a sequential $Z^\prime$ gauge boson is excluded at 95$\%$ C.L. for masses below 2.39 TeV in the electron channel, 2.19 TeV in the muon channel, and 2.49 TeV in the two channels combined~\cite{Zpatlas-mass}; $Z^\prime$ bosons coming from $E_6$-motivated models are excluded at 95$\%$ C.L. for masses below 2.09-2.24 TeV~\cite{Zpatlas-mass}. In accordance with CMS results, in the context of the sequential $Z$ model and the superstring-inspired model, the lower mass limits at 95$\%$ C.L. for the $Z^\prime$ gauge boson correspond to 2.59 TeV and 2.26 TeV, respectively~\cite{Zpcms-mass}. Motivated by the above results, we will take a lower mass limit for the $Z_H$ gauge boson to be 2.6 TeV in order to explore the physical possibilities for the $Z_H \to \gamma H$ decay. In a previous work it has been studied the dominant decays of the $Z_H$ boson~\cite{tesis} in the context of the LTHM. We employ this information to compute the total decay width of the $Z_H$ boson for different values of the energy scale $f$ proposed above.

\begin{figure}[htb!]
\centering
\subfigure[]{\includegraphics[scale=0.66]{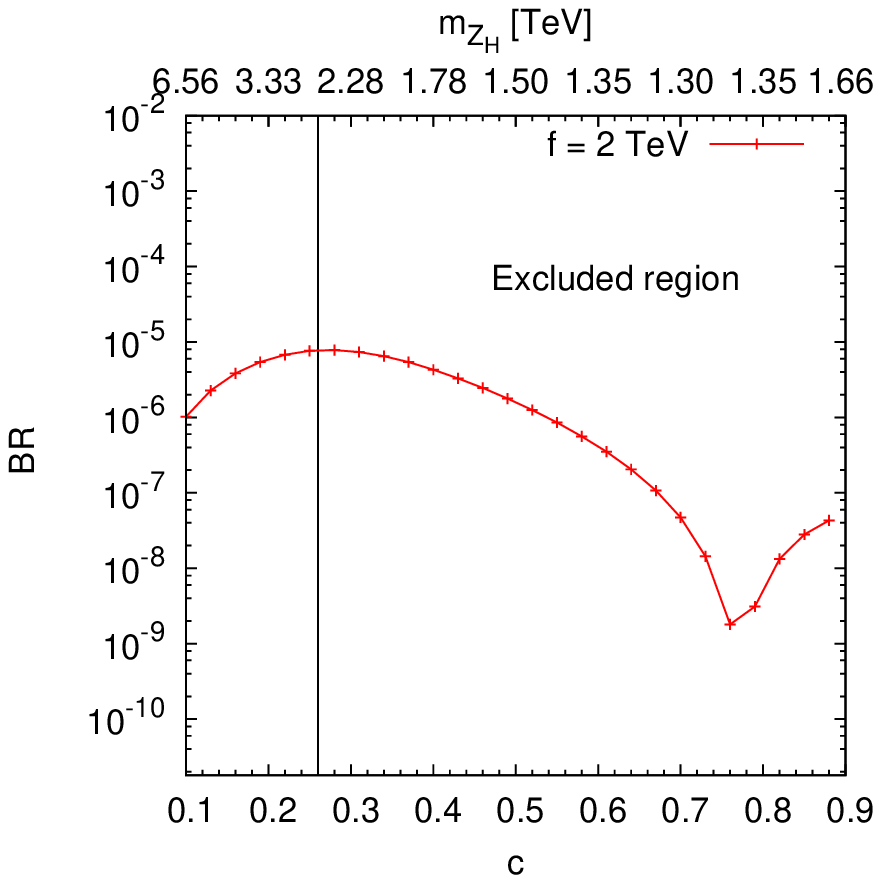}}\;
\subfigure[]{\includegraphics[scale=0.66]{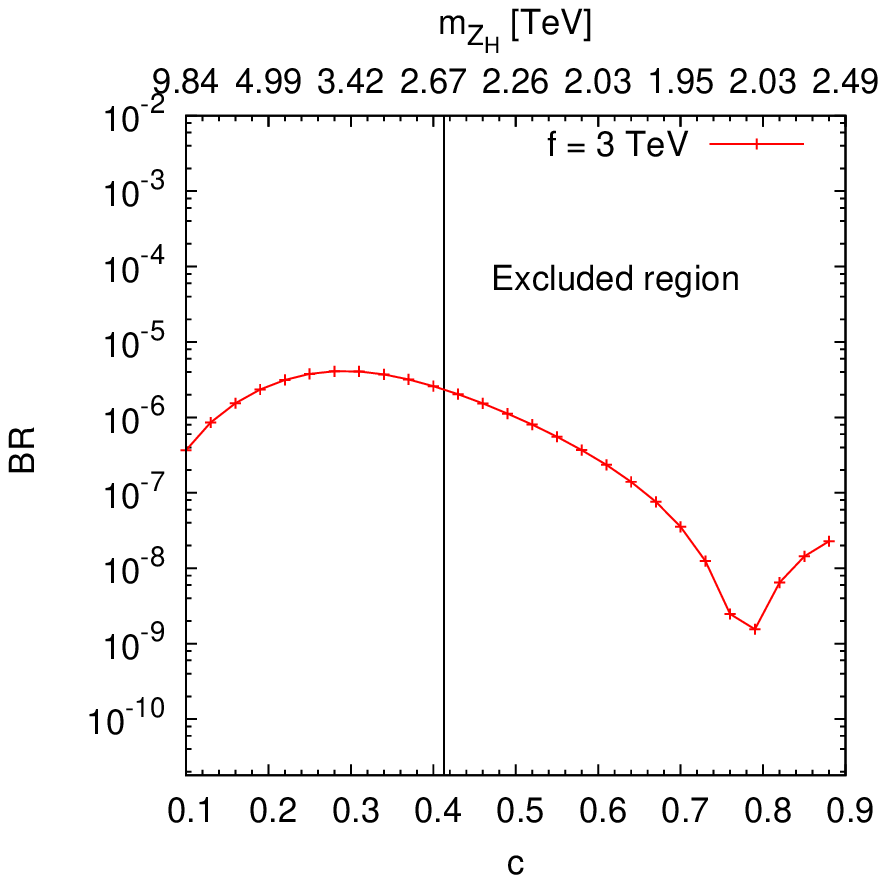}}
\subfigure[]{\includegraphics[scale=0.66]{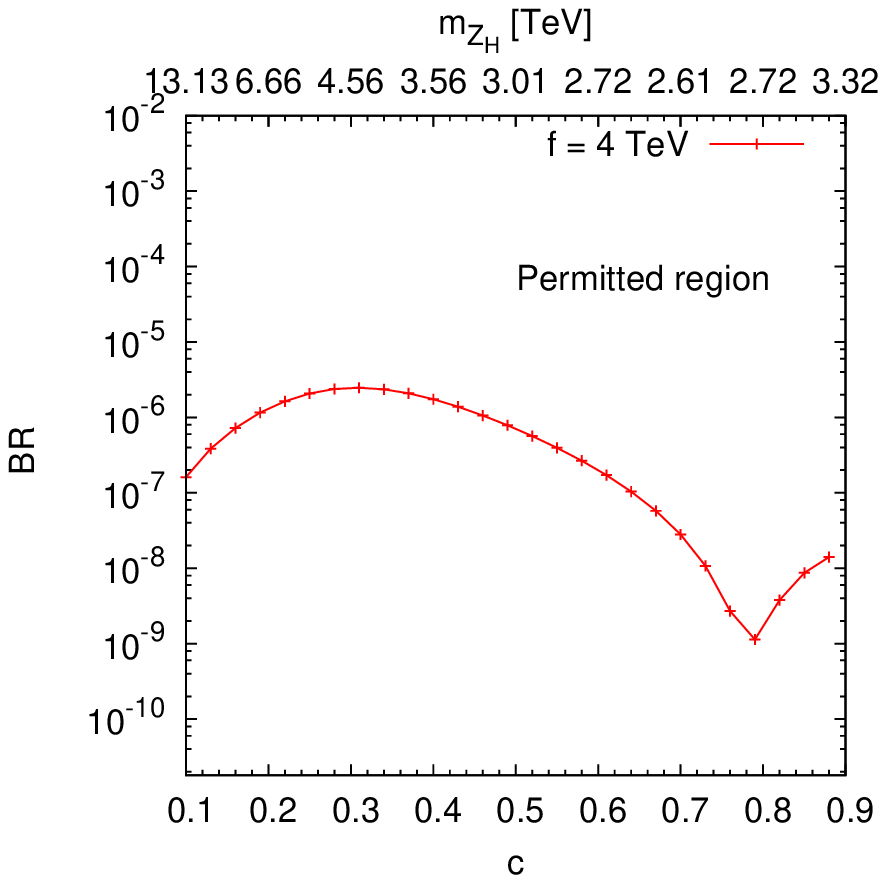}}
\caption{Branching ratio for the $Z_H\to \gamma H$ decay as a function of $c$ parameter and $m_{Z_H}$. (a) For $f=2$ TeV. (b) For $f=3$ TeV. (c) For $f=4$ TeV.}
\label{BRc}
\end{figure}

From Fig.~\ref{BRc}(a), for $f=2\,\,\mathrm{TeV}$, it can be observed that the permitted region corresponds to $0.1 < c < 0.26$ for a $Z_H$ mass interval $6.56\,\,\mathrm{TeV}>m_{Z_H}>2.6\,\,\mathrm{TeV}$, where the branching ratio ranges from $1.02\times10^{-6}$ to $7.77\times10^{-6}$; the maximum value of the branching ratio is $7.77\times10^{-6}$ for $c=0.26$ and $m_{Z_H}=2.68$ TeV. In Fig.~\ref{BRc}(b), for $f=3\,\,\mathrm{TeV}$, we can observe a permitted region for the $c$ parameter ranging between $0.1 < c < 0.41$, due to $Z_H$ mass interval $9.84\,\,\mathrm{TeV}>m_{Z_H}>2.6\,\,\mathrm{TeV}$, being the associated branching ratio around $3.67\times10^{-7}\,\textendash\,2.4\times10^{-6}$; in this case, the maximum value of the branching ratio is $4.12\times10^{-6}$ for $c=0.29$ and $m_{Z_H}=3.53$ TeV. Finally in Fig.~\ref{BRc}(c), we show the branching ratio as a function of $c$ and $m_{Z_H}$ for $f=4$ TeV. This figure tell us that the whole interval for the $c$ parameter $0.1 < c < 0.9$ is permitted, accordingly with the interval $2.6\,\,\mathrm{TeV}<m_{Z_H}<13.13\,\,\mathrm{TeV}$, where the related branching ratio is as high as $2.47\times10^{-6}$ for $c=0.31$ and $m_{Z_H}=4.43$ TeV. Even when predicted branching ratio is small (of the order of $10^{-5}$), the LHC luminosity at final stage of operation (14 TeV at the center of mass energy) is expected to be around 3000 $\mathrm{fb}^{-1}$~\cite{final-luminosity} which could counteract this situation. In fact, to explore the predictability of the LTHM, let us consider different values of the $Z_H$ mass for which it is produced few events. For $f=2$ TeV and around $c=0.19$ or equivalently $m_{Z_H}=3.5$ TeV, we estimate 2 events. For $f=3$ TeV and $c=0.28$ or $m_{Z_H}=3.64$ TeV, we found around 1 event. For $f=4$ TeV and $c=0.4$ or $m_{Z_H}=3.56$ TeV, we calculated less than 1 event. Moreover, the maximum number of events estimated for $f=2,3,4$ TeV are 16 ($c=0.26, m_{Z_H}=2.6$ TeV), 4 ($c=0.41, m_{Z_H}=2.62$ TeV) and 0.76 ($c=0.45, m_{Z_H}=3.25$ TeV), respectively.

\subsection{Background estimation for the $\gamma H$ production in LHC}

\begin{figure}[htb!]
\begin{center}
\includegraphics[scale=0.5]{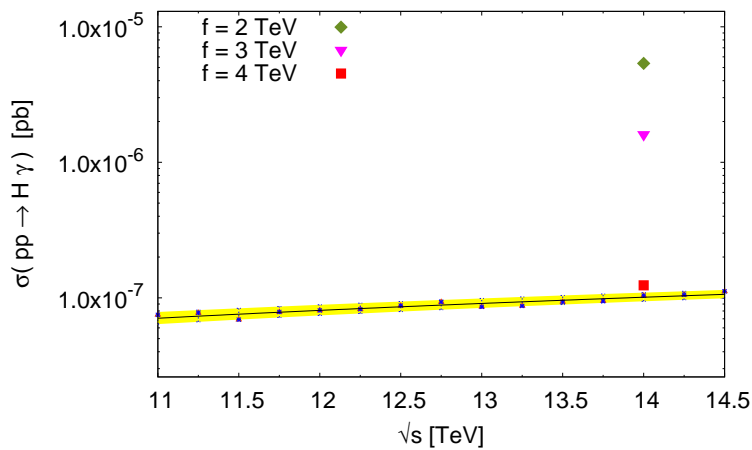}
\caption{SM-background cross section for the Higgs-photon associated production at LHC as a function of center of mass energy. The various dots represent $\sigma(pp\to Z_H X\to \gamma H)$ at $\sqrt{s}=14$ TeV for $m_{Z_H}=2.6$ TeV.}\label{SM-background}
\end{center}
\end{figure}

At the LHC, the Higgs boson production in association with a photon may be produced from quark-antiquark annihilation. In gluon-gluon annihilation, the Higgs-photon associated production is forbidden by Furry's theorem because of properties of color singlet state of gluons~\cite{Abbasabadi:1997zr}. Different results were reported where the $H \gamma$ final states are produced in $pp$ collisions via $\bar{d} + d \to H \gamma$ subprocess or via weak-boson fusion and  $e^+ e^-$ colliders \cite{Passarino:2013nka,Arnold:2010dx,Hu:2014eia}. For the process $\bar{d} + d \to H \gamma$ the cross section was obtained at $\sqrt{s}=8$ TeV using values of transverse momentum, $p_T$, ranging from $30$ GeV to $300$ GeV. For our analysis we take $pp$ collisions with a center-of-mass energy of $14$ TeV and introduce kinematical cuts based on experimental values employed by ATLAS Collaboration for the transverse momentum $30\;\; \mathrm{GeV}<p_T<150\;\; \mathrm{GeV}$ and transverse energy $E_T$ from $300$ GeV to $1000$ GeV for the photon and pseudorapity region $|\eta|< 1.37$~\cite{Aad:2010sp,Aad:2013zba}. Taking in account this, we simulated the $pp \to H \gamma$ process corresponding to SM background by using Whizard Event Generator package \cite{Kilian:2007gr} along with parton distribution functions (PDFs) CTEQ5, CTEQ6~\cite{CTEQ6} and MSTW2008NLO~\cite{MSTW2008NLO}. We note that the PDFs used produce similar results for the production cross section in question. In addition, we would like to mention that with the used cuts, the SM background results (see Fig.~\ref{SM-background}) are suppressed in comparison with the corresponding LTHM results. Concretely, at center of mass energy of 14 TeV we found a SM background cross section of $1.077\times10^{-7}$ pb (for CTEQ5), while in the less optimistic scenario, our calculated cross section via LTHM is $2.19\times10^{-7}$ pb (for $f=4$ TeV and $m_{Z_H}=3.56$ TeV). For $f=2$ ($f=3$) TeV and $m_{Z_H}=3.5$ ($m_{Z_H}=3.32$) TeV respectively, we obtain a LTHM production cross section $\sigma(pp\to \gamma H)$ of the order of $10^{-6}$ pb.

To distinguish the LTHM signal from SM background we compute the Higgs gamma Drell-Yan production (background) at 95\% of confidence level, where systematic uncertainties has been neglected (see Fig.~\ref{SM-background}). From this figure, we observe that all the points corresponding to $pp\to Z_H X\to \gamma H$ process at 14 TeV are above the confidence level band of SM background. These LTHM results could be attractive for the search of new physics, however, they must be taken carefully since we have not considered their respective error bars. Moreover, it may be recalled that the purpose of this work relies on the LTHM predictability for the physical parameter space~\cite{Reuter} by using the $Z_H\to \gamma H$ decay, to this end we present at the most a thorough estimation.

Finally, it is essential to point out that it has been shown the important influence of K factors on QCD NLO corrections to Drell-Yan production in both Tevatron and LHC~\cite{Altarelli}, where it is found that corrections to the one-loop level under certain scenarios are important. In this sense, one of the goals of this work has been calculating the $Z_H$ production cross section at the LHC via $Z_H$ Drell-Yan production, which has been made possible through the implementation of PDFs such as MSTW2008NLO~\cite{MSTW2008NLO} in The Whizard event generator. These PDFs already take into account corrections for NLO Drell-Yan production at the LHC~\cite{PDF,CTEQ6,MSTW2008NLO}. Indeed, as already mentioned before, while for the $Z_H$ production and the SM background analysis it has been used the CTEQ5 PDF, we stress that the numerical simulation also was performed using MSTW2008NLO and CTEQ6 and no significant changes between the three PDFs are appreciated, as the results provided by these PDFs hardly vary even in the most significant digit. For consistency, we use CTEQ5 in accordance with previous experimental analyses that consider this PDF for studies of $Z'$ Drell-Yan production~\cite{PDF-EXP}.

\section{Conclusions}
\label{CON}

The LTHM resides on a nonlinear sigma model with a $SU(5)$ global symmetry and the gauged subgroup $[SU(2)_1\otimes U(1)_1]\otimes [SU(2)_2\otimes U(1)_2]$, where it is predicted the existence of heavy gauge bosons, particularly, a new neutral massive boson known as $Z_H$. This gauge boson is another $Z^\prime$ type gauge boson which at present is under experimental scrutiny at the LHC. Although the parameters space of the LTHM has been severely constrained, yet there is room left to test the predictability of the model. In specific, the $Z_H\to \gamma H$ decay was used to explore the current parameters space of the LTHM, where we have analyzed physical regions according with experimental bounds and results; specifically we have taken the following parameters: $f=2,3,4$ TeV for $0.1<c<0.9$. It is found that for $f=2$ TeV there is a permitted region $0.1<c<0.26$ corresponding to $6.56\,\,\mathrm{TeV}>m_{Z_H}>2.6\,\,\mathrm{TeV}$. In particular, for a $m_{Z_H}=3.5$ TeV it is calculated around 2 events for the $Z_H\to \gamma H$ decay at LHC operating at $14$ TeV. Similarly, for $f=3$ TeV the permitted region is $0.1 < c < 0.41$ or a $Z_H$ mass interval $9.84\,\,\mathrm{TeV}>m_{Z_H}>2.6\,\,\mathrm{TeV}$. In this case, for $m_{Z_H}=3.64$ TeV it is estimated 1 event for the process in question. Finally, for the same process and taking $f=4$ TeV we have found the permitted region in the interval of masses $2.6\,\,\mathrm{TeV}<m_{Z_H}<13.13\,\,\mathrm{TeV}$. Here, it is computed less than 1 event for $m_{Z_H}=3.56$ TeV. Although we have chosen specific values of $m_{Z_H}$ to get few events, our numerical results tell us that there are several intervals in which the number of events are larger than the previous ones. For instance, for $f=2$ and $c=0.26$ we could obtain tens of events for the $Z_H\to \gamma H$ decay. To explore future experimental possibilities of the decay in question, we have performed a careful estimation of the SM background for the $pp\to H\gamma$ reaction, by using the CTEQ5, CTEQ6, and MSTW2008NLO PDFs included in the WHIZARD event generator. Taking into account current experimental kinematical cuts, the SM-background cross section results below the computed LTHM cross section for the $pp\to Z_H X\to H\gamma$.

\section*{Acknowledgments}
This work has been partially supported by CONACYT and CIC-UMSNH. I. Cort\'es-Maldonado thanks to J. J. Toscano for fruitful discussions.\\

\end{document}